\documentstyle[12pt,epsfig]{article}
\def\unit{\hbox to 3.3pt{\hskip1.3pt \vrule height 7pt width .4pt \hskip.7pt
\vrule height 7.85pt width .4pt \kern-2.4pt
\hrulefill \kern-3pt
\raise 4pt\hbox{\char'40}}}

\newcommand{\gtrsim}
{\ \rlap{\raise 2pt\hbox{$>$}}{\lower 2pt \hbox{$\sim$}}\ }
\newcommand{\lessim}
{\ \rlap{\raise 2pt\hbox{$<$}}{\lower 2pt \hbox{$\sim$}}\ }
 
\newcommand{\ea}{{ et al.}}

\def\bsl{BR_{SL}} 
\def\bsg{b\to s g}
\def\bsf{B\to X_s \gamma}
\def\bfi{B\to X_s \phi}
\def\beq{\begin{equation}}
\def\eeq{\end{equation}}
\def\bea{\begin{eqnarray}}
\def\eea{\end{eqnarray}}
\def\ba{\begin{array}}
\def\ea{\end{array}}
\def\mg{m_{\tilde g}}
\def\tm{{\tilde m}}
\def\bs{{\bar s}}
\def\gappeq{\mathrel{\rlap {\raise.5ex\hbox{$>$}}
{\lower.5ex\hbox{$\sim$}}}}
\def\permil{$\%\raise.20ex\hbox{$_0$}}
\def\lappeq{\mathrel{\rlap{\raise.5ex\hbox{$<$}}
{\lower.5ex\hbox{$\sim$}}}}

\def\pl#1#2#3{{\it Phys. Lett. }{\bf B#1~}(19#2)~#3}
\def\zp#1#2#3{{\it Z. Phys. }{\bf C#1~}(19#2)~#3}
\def\prl#1#2#3{{\it Phys. Rev. Lett. }{\bf #1~}(19#2)~#3}

\def\pr#1#2#3{{\it Phys. Rev. }{\bf D#1~}(19#2)~#3}
\def\np#1#2#3{{\it Nucl. Phys. }{\bf B#1~}(19#2)~#3}

\begin{document}
\topmargin -1.0cm
\oddsidemargin -0.8cm
\evensidemargin -0.8cm
\pagestyle{empty}
\begin{flushright}
CERN-TH/96-73
\end{flushright}
\vspace*{5mm}
\begin{center}
{\Large \bf The Chromomagnetic Dipole Operator\\
and the $B$ Semileptonic Branching Ratio}\\
\vspace{2.5cm}
{\large M.~Ciuchini\footnote{On leave of absence from INFN, Sezione
Sanit\`a, Rome, Italy.}}\\
{\small Theory Division, CERN, Geneva, Switzerland}\\
\vspace{0.5cm}
{\large E.~Gabrielli}\\
{\small INFN, Sezione di Roma II, Rome, Italy}\\
\vspace{0.5cm}
{\large G.F.~Giudice\footnote{On leave of absence from INFN, Sezione di
Padova, Padua, Italy.}}\\
{\small Theory Division, CERN, Geneva, Switzerland}\\
\vspace*{2cm}
ABSTRACT
\end{center}

We consider the possibility of having a large branching ratio for the
decay $b\to s g$ coming from an enhanced Wilson coefficient of the
chromomagnetic dipole operator. We show that values of $BR(b\to s g)$
up to $\sim 10\%$ or more are
compatible with the constraints coming from the CLEO
experimental results on $BR(B\to X_s\gamma)$ and $BR(B\to X_s\phi)$.
Such large values can reconcile
the predictions of both the semileptonic branching ratio and the charm
counting with the present experimental results. We also discuss a
supersymmetric model with gluino-mediated flavour violations, which
can account for such large values of $BR(b\to s g)$.

\vfill
\begin{flushleft}
CERN-TH/96-73\\
April 1996
\end{flushleft}
\eject
\pagestyle{empty}
\setcounter{page}{1}
\setcounter{footnote}{0}

\baselineskip20pt
\pagestyle{plain}

\section{Introduction}
The world average of all measurements of the $B$-meson semileptonic 
branching ratio is \cite{brsl}
\beq
BR_{SL}^{exp}\equiv BR(B\to X e \bar \nu_e)=(10.4\pm 0.4)\% ~.
\label{eq:brslexp}
\eeq
This result is considerably smaller than the theoretical prediction
in the parton model, where $\bsl \sim 13$--$15 \%$ \cite{alt}.
Furthermore
$1/m_Q^2$ non-perturbative corrections cannot reduce the predicted
$\bsl$ below $12.5 \%$ \cite{big}.
However it has recently been found that charm mass
corrections to $\Gamma (b\to c\bar c s)$ are large and can
further reduce the theoretical prediction for $\bsl$ \cite{bbbg}:
\beq
\bsl =\left\{ \begin{array}{ll}
(12.0\pm 0.7\pm 0.5^{+0.9}_{-1.2}\pm 0.2) \% & \mbox{on-shell scheme}\\
(11.3\pm 0.6\pm 0.7^{+0.9}_{-1.7}\pm 0.2) \% & \overline{\mbox{MS}}~
\mbox{scheme}\\
\end{array} \right.
\label{bal}
\eeq
Here the errors correspond to the uncertainties on $m_b$, $\alpha_s(M_Z)$, 
the renormalization scale $\mu$ and the other parameters present in the
calculation.
It may now seem that the discrepancy in the $B$ semileptonic branching ratio
has essentially disappeared.
However,
as we lower the theoretical prediction for $\bsl$ by increasing
$\Gamma (b\to c\bar c s)$, we simultaneously increase the prediction for
$n_c$, the average number of charm hadrons produced per $B$ decay. 
Indeed the theoretical prediction for $n_c$ obtained from eq.~(\ref{bal}) is
\beq
n_c =\left\{ \begin{array}{ll}
1.24\pm 0.05\pm 0.01 & \mbox{on-shell scheme}\\
1.30\pm 0.03\pm 0.03\pm 0.01 & \overline{\mbox{MS}}~\mbox{scheme}\\
\end{array} \right.
\label{nc}
\eeq
where the first error corresponds to the quark mass uncertainties, the
second to the $\alpha_s$ variation in $\overline{\mbox{MS}}$ and the last
to all the other uncertainties. This prediction is larger than the
present world average of the $n_c$ measurements \cite{brsl}
\beq
n_c^{exp}=1.117\pm 0.046~.
\label{eq:ncexp}
\eeq
The real
problem is therefore to understand the presence of a low $\bsl$ together
with a low value of $n_c$. The theoretical difficulty is summarized in
fig.~\ref{fig:brslnc}.

It could well be that this is just an experimental problem; measurements
of $n_c$ rely on model assumptions about charm hadron production which
have been questioned in ref.~\cite{que}. 
Recently it has been proposed \cite{nes}
that ``spectator effects" can be at the origin of the problem. Although
these effects cannot be fully computed without a knowledge of the hadronic
parameters, potentially they can bring $\bsl$ close to the experimental
value. They also result in a simultaneous, although modest,
decrease of $n_c$ \cite{nes}. On the other hand, if ``spectator effects"
are responsible for the observed low value of the ratio $\tau (\Lambda_b)
/\tau(B_d)=0.76\pm 0.05$, they then tend to increase the prediction for
$\bsl$, making the problem even more pressing.
Recently it has also been suggested \cite{ampr} that the
data indicate the presence of $1/m_Q$ corrections in non-leptonic
decays, which are not present in HQET. This has been attributed to a
possible failure of the operator product expansion near the physical
cut and a corresponding violation of the local quark--hadron duality.
In particular, using the phenomenological recipe of
replacing the quark mass by the decaying hadron mass in the $m^5$
factor in front of all non-leptonic widths, $\bsl$, $n_c$, and
$\tau (\Lambda_b)/\tau(B_d)$ can be reconciled with their measured values
\cite{ampr}.

In this paper we discuss the effect of new physics on $\bsl$ and $n_c$.
The possibility that large contributions to $\Gamma (\bsg )$ can eliminate
the disagreement between experiments and Standard Model predictions was
first proposed in ref. \cite{bh}, and then studied in much further detail in
ref. \cite{kag}. Our goal here is to refine previous analyses and consider new
constraints.

In sect.~\ref{sec:modind} we perform a model-independent analysis of
$\Gamma (\bsg )$
and the possible constraints coming from $\Gamma (\bsf )$ and
$\Gamma(\bfi )$, extending the results of ref. \cite{kag}.
In particular we show that $\Gamma (\bfi )$, in spite
of its sensitivity to penguin operators, provides a poor probe of New
Physics effects. In sect.~\ref{sec:model} we discuss a
supersymmetric model with
gluino-mediated flavour violations, first suggested by Kagan \cite{kag},
which can explain a large enhancement of $\Gamma (\bsg )$.

\section{Model-Independent Analysis}
\label{sec:modind}

The effective Hamiltonian relevant for $\Delta B=1$ decays is given
by \cite{bjlw}
\begin{eqnarray}
H_{eff}^{\Delta B=1}&=&\frac{G_F}{\sqrt{2}}\Biggl[ V^*_{cb}V_{cs} \left(
C_1(\mu) Q_1^c(\mu)+C_2(\mu) Q_2^c(\mu)\right)+\\
&&  V^*_{ub}V_{us}\left(C_1(\mu) Q_1^u(\mu)+C_2(\mu) Q_2^u(\mu)\right)-
V^*_{tb}V_{ts}\sum_{i=3}^{12} C_i(\mu) Q_i(\mu)\Biggr]\,,\nonumber
\end{eqnarray}
where $V_{ij}$ are the CKM matrix elements and $C_i(\mu)$ are the Wilson
coefficients evaluated at the scale $\mu$ of order $m_b$.
The dimension-six local operator basis $Q_i$ is
given by
\begin{eqnarray}
Q_1^{q}&=&(\bar s_\alpha q_\beta)_{V-A}
(\bar q_\beta b_\alpha)_{V-A}\nonumber\\
Q_2^{q}&=&(\bar s_\alpha q_\alpha)_{V-A}
(\bar q_\beta b_\beta)_{V-A}\nonumber\\
Q_{3,5}&=&(\bar s_\alpha b_\alpha)_{V-A}\sum_q
(\bar q_\beta q_\beta)_{V\mp A}\nonumber\\
Q_{4,6}&=&(\bar s_\alpha b_\beta)_{V-A}\sum_q
(\bar q_\beta q_\alpha)_{V\mp A}\nonumber\\
Q_{7,9}&=&\frac{3}{2}(\bar s_\alpha b_\alpha)_{V-A}\sum_q
e_q (\bar q_\beta q_\beta)_{V\pm A}\nonumber\\
Q_{8,10}&=&\frac{3}{2}(\bar s_\alpha b_\beta)_{V-A}\sum_q
e_q (\bar q_\beta q_\alpha)_{V\pm A}\nonumber\\
Q_{11}&=&\frac{g_s}{16\pi^2}m_b \bar s_\alpha \sigma^{\mu\nu}_{V+A}
t^A_{\alpha\beta} b_\beta G^A_{\mu\nu}\nonumber\\
Q_{12}&=&\frac{e}{16\pi^2}m_b \bar s_\alpha \sigma^{\mu\nu}_{V+A} b_\alpha
F_{\mu\nu}.
\end{eqnarray}
Here $(\bar q_1 q_2)(\bar q_3 q_4)$ denotes current--current products,
$V\pm A$ indicates the chiral structure and $\alpha ,\, \beta$ are colour
indices. Moreover, $g_s$ ($e$) is the strong (electromagnetic) coupling
constant, $G^A_{\mu\nu}$ ($F_{\mu\nu}$) is the gluon (photon) field
strength, and the $t^A$ are the $SU(N)$ colour matrices normalized so that
$Tr(t^A t^B)=\delta^{AB}/2$.

This Hamiltonian is known at the next-to-leading order (NLO), as far as one
separately considers the
current--current, gluon and photon penguin operators $Q_1$-$Q_{10}$ on the one
hand \cite{bjl} and the magnetic-type
operators $Q_{11}$-$Q_{12}$ on the other hand \cite{mm1}. Unfortunately
the mixing between these two classes of operators is at present known only at
the leading order (LO) \cite{cfmrs2}.

\begin{table}[t]
\begin{center}
\begin{tabular}{|c|c|}
\hline
Parameter & Value \\ \hline\hline
$\vert V_{ts}^* V_{tb} \vert^2/\vert V_{cb}\vert^2$ & $0.95$ \\ \hline
$\alpha_s(M_Z)$ & 0.117 \\ \hline
$m_t$ (GeV) & $174$ \\ \hline
$m_b$ (GeV) & $5$ \\ \hline
$m_s$ (MeV) & 500 \\ \hline
$m_c/m_b$ & $0.316$ \\ \hline
$M_\phi$ (GeV) & 1.019 \\ \hline
$g_\phi^2$ $(\mbox{GeV}^4)$ & 0.0586 \\ \hline
\end{tabular}
\caption[]{ \it{Central values of the input parameters used in the analysis}}
\label{tab:const}
\end{center}
\end{table}

The branching ratio of the decay $b\to s g$ is given by \cite{cfmrs3}
\begin{equation}
BR(b\to s g)=BR_{SL}^{exp} \frac{\vert V_{ts}^* V_{tb}\vert^2}
{\vert V_{cb}\vert^2}\frac{2\alpha_s(\mu)}{\pi g(m_c/m_b)}\vert C_{11}^{eff}
(\mu)\vert^2\,,
\label{eq:bsg}
\end{equation}
where $BR_{SL}^{exp}$ is the measured semileptonic branching ratio, the
phase-space factor $g(z)$ is given in the Appendix, $\mu=O(m_b)$ and
$C_{11}^{eff}$ is the
renormalization-scheme invariant coefficient\footnote{Using
calculations in the 't Hooft--Veltman
scheme, as we do, $C_{11,12}^{eff}=C_{11,12}^{HV}$. Therefore we denote them
$C_{11,12}$ in the following.} introduced
in ref. \cite{bmmp}. With the input values given in table~\ref{tab:const},
the LO Standard Model prediction is
\beq
BR^{SM}(\bsg )=(2.3\pm 0.6)\times 10^{-3}~,
\eeq
where the error is mainly due to the variation of $\mu$ between $m_b/2$ and
$2m_b$ and to the uncertainty on $\alpha_s(M_Z)$. 
Inclusion of the known part of the NLO corrections reduces
the central value and the $\mu$ dependence, but introduces a significant
scheme dependence \cite{cfmrs3}.

In the effective Hamiltonian
approach, physics beyond
the weak scale affects only the initial conditions of the Wilson
coefficients, namely $C_i(M_W)$. Therefore these coefficients
can be used to parametrize new physics effects without
referring to a specific model. 
In the case at hand, we assume that the initial condition of the
Wilson coefficient $C_{11}(M_W)$ is an independent variable and define
\beq
r_g=\frac{C_{11}(M_W)}{C_{11}^{SM}(M_W)}~,
\label{rg}
\eeq
where
\bea
C_{11}^{SM}(M_W)&=&g_1(m_t^2/m_W^2)~,\\
g_1(x)&=&\frac{x(-x^2+5x+2)}{4(x-1)^3}-\frac{3x^2\log x}{2(x-1)^4}~.
\nonumber
\eea
In terms of $r_g$, the branching ratio of $\bsg$ is given by
\beq
BR(\bsg )=2.73\times 10^{-2} (0.15+0.14~r_g)^2~,
\eeq
assuming central values for all the relevant parameters.

\begin{figure}[t]
\centering
\epsfxsize=0.9 \textwidth
\leavevmode\epsffile{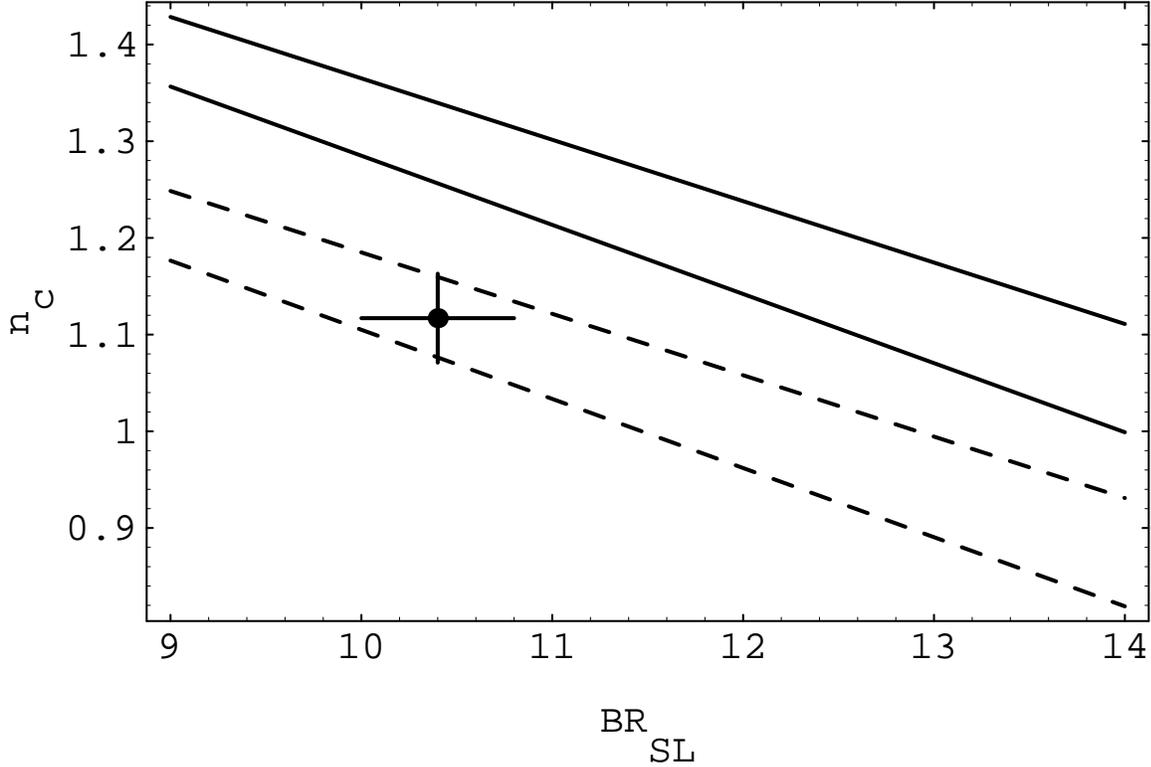}
\caption[]{\it{Correlation between the semileptonic branching ratio
$BR_{SL}$ and the number of charms per $B$ decay $n_c$. The solid band
is the Standard Model prediction, including the theoretical uncertainties,
while the dashed one is obtained assuming a $BR(b\to s g)=9\%$. The
experimental data point is also shown.}}
\label{fig:brslnc}
\end{figure}

An enhanced value of $BR(\bsg )$ affects the relation between $n_c$ and 
$\bsl$. We find
\begin{equation}
n_c=2-(2+R_\tau+R_{ud}+2R_{c\!\!/})BR_{SL}-2 BR(b\to s g)~,
\label{eq:nc}
\end{equation}
where \cite{bdy}
\begin{eqnarray}
R_\tau&=&\frac{\Gamma(b\to c \tau \nu)}{\Gamma(b\to c e \nu)}=0.25\nonumber\\
R_{ud}&=&\frac{\Gamma(b\to c \bar u d')}{\Gamma(b\to c e \nu)}=4.0\pm 0.4
\nonumber\\
R_{c\!\!/}&=&\frac{\Gamma(b\to\mbox{no charm})}{\Gamma(b\to c e \nu)}=0.25\pm
0.10.
\end{eqnarray}
Here $\Gamma(b\to\mbox{no charm})$ is the sum of all $B$ decay widths into
charmless final states different from $sg$. Notice that in eq.~(\ref{eq:nc})
we have eliminated the dependence on $\Gamma (b\to c\bar c s)$, which is
the main source of theoretical uncertainty. If we require that the
experimental central values of $\bsl$ and $n_c$, eqs.~(\ref{eq:brslexp})
and (\ref{eq:ncexp}), lie within the theoretical
uncertainty, we need
$7\%<BR(\bsg )<11\%$. This corresponds to $10.2<r_g<13.1$ or
$-12.3<r_g<-15.2$. For instance the
effect of $BR(\bsg )= 9\%$ in eq. (\ref{eq:nc}) is shown in
fig.~\ref{fig:brslnc} (see the dashed band).

We now want to perform a model-independent analysis of the constraints 
on $r_g$. The first constraint comes from the observation of the 
inclusive decay $\bsf$. The LO expression for $BR(\bsf )$
is \cite{cfmrs3,bmmp}
\begin{equation}
BR(B\to X_s\gamma)=BR_{SL}^{exp} \frac{\vert V_{ts}^* V_{tb}\vert^2}
{\vert V_{cb}\vert^2}\frac{3\alpha_e}{2 \pi g(m_c/m_b)}\vert C_{12}^{eff}(\mu)
\vert^2.
\label{eq:bsgamma}
\end{equation}
It is well known that this
branching ratio has very large QCD corrections. It also has a significant
theoretical
uncertainty, mainly due to the scale dependence in the Wilson coefficient,
to be taken into account. Also in this case, the known NLO terms decrease the
central value and the $\mu$ dependence, but introduce a sizeable scheme
dependence, to be cancelled by the unknown terms.

\begin{figure}[t]
\centering
\epsfxsize=0.8 \textwidth
\leavevmode\epsffile{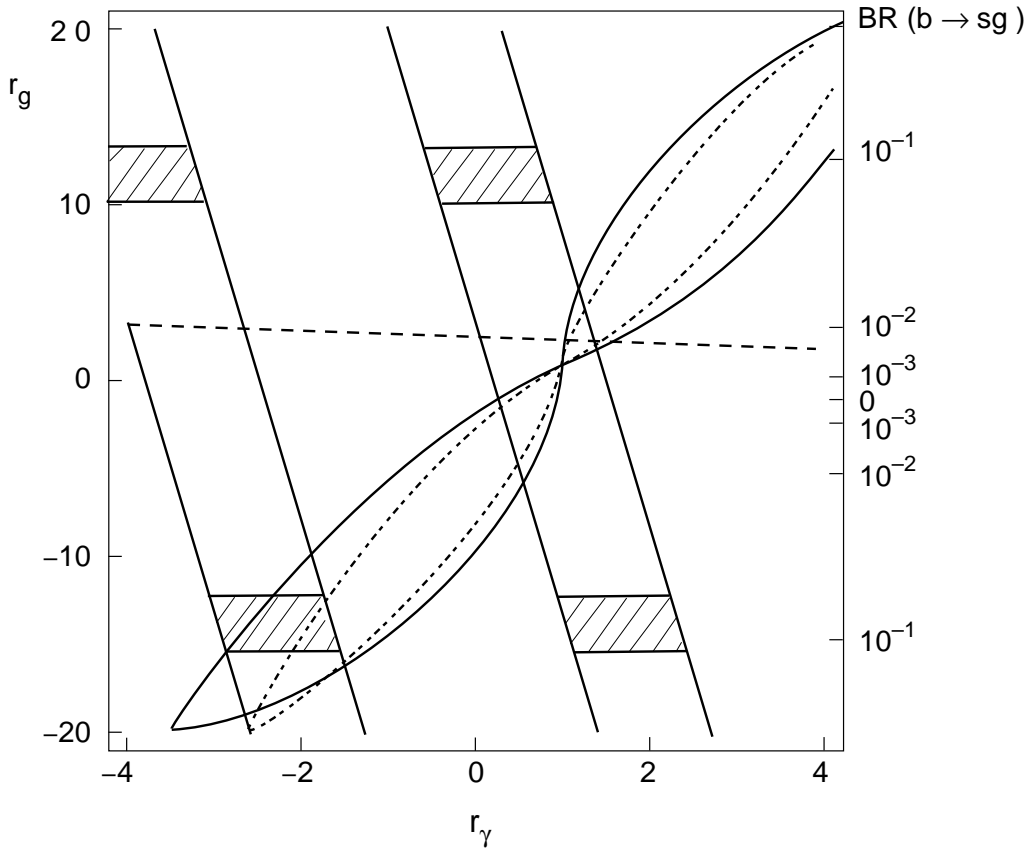}
\caption[]{\it{Correlation between the ratios $r_g$
and $r_\gamma$, defined in eqs.~(\ref{rg}) and (\ref{rga}). The scale on the
right shows the corresponding values of $BR(b\to s g)$. The regions outside
the two oblique bands are excluded by the measured $BR(B\to X_s\gamma)$. The
region above the dashed line is excluded by the upper limit on the
$BR(B\to X_s\phi)$, assuming $\xi=1/3$. However, the theoretical error is large
and not completely under control in this case, so that the limit is plotted
only for illustrative purpose. Finally the two oblated regions delimit the
ranges allowed in the supersymmetric model with LL (dotted line) and
LR (solid line) flavour mixing obtained for $\mg > 200$ GeV and
$\tm > 100$ GeV.}}
\label{fig:r11r12}
\end{figure}

The dependence on $r_g$ in eq.~(\ref{eq:bsgamma})
comes from the operator mixing in the QCD 
renormalization group equations. Analogously to eq.~(\ref{rg}), we define
\beq
r_\gamma =\frac{C_{12}(M_W)}{C_{12}^{SM}(M_W)}~,
\label{rga}
\eeq
where
\bea
C_{12}^{SM}(M_W)&=&f_1(m_t^2/m_W^2)~,\\
f_1(x)&=&\frac{x(-8x^2-5x+7)}{12(x-1)^3}+\frac{x^2(3x-2)\log x}{2(x-1)^4}~.
\nonumber
\eea
We can now rewrite 
eq. (\ref{eq:bsgamma}) in terms of
$r_g$ and $r_\gamma$ as
\begin{equation}
BR(B\to X_s\gamma)=7.10\times 10^{-4} (0.017\, r_g+0.273\, r_\gamma+0.313)^2,
\label{eq:bsgamma1}
\end{equation}
where the theoretical error is not shown.
This formula gives a model-independent parametrization of $BR(B\to X_s\gamma)$. 
Figure~\ref{fig:r11r12} shows the ranges of $r_g$ and $r_\gamma$ allowed by the 
measurement $BR(\bsf )=(2.32\pm 0.67)\times 10^{-4}$ \cite{bsgamma}.
The existence of two different 
bands corresponds to the overall sign ambiguity of $C_{12}^{eff}(\mu )$
in eq.~(\ref{eq:bsgamma}).
The size of the bands accounts for both the experimental error at the
$1\sigma$ level and a (linearly added) theoretical error of $30\%$
\cite{bmmp}.
Figure \ref{fig:r11r12} shows that very large
enhancements of $C_{11}(M_W)$ are allowed, since they can be compensated by
appropriate, and relatively small, changes of $C_{12}(M_W)$.
A large values of $C_{11}(M_W)$
can result in a modification of the photon energy spectrum in $\bsf$
and could be tested in the future if the threshold in the observed $E_\gamma$
is appropriately lowered\footnote{We thank A. Ali for pointing out this
possible experimental test.} \cite{ali}.
The solution of the $\bsl$--$n_c$ problem in terms of $\Gamma (b\to s g)$
singles out the four disconnected regions in the $r_g$-$r_\gamma$ plane
shown in fig.~\ref{fig:r11r12}.

The extraction of analogous constraints from the measured upper limit
\cite{bsphi}
\beq
BR(B\to X_s\phi) < (1.1-2.2)\times 10^{-4}\qquad (90\%~\mbox{CL})
\eeq
is more involved. Available theoretical
estimates \cite{dh,dht} rely on some assumptions which are
questionable. In particular, hadronic matrix elements are calculated by
combining perturbation theory with the factorization method. Moreover they
made a strong assumption on the quark momenta distribution
inside the $\phi$ meson ($p_s=p_{\bar s}=p_\phi/2$). These
hypotheses result in a pure two-body decay $b\to s\phi$, while for example
string fragmentation models
indicate high multiplicity final states for the $b\to sg$ decay \cite{sch}. The
same assumptions are also used to extract the experimental limit from the
measurements. Thus, constraints coming
from this limit have large theoretical uncertainties and should be
taken with caution.

Nonetheless, in the following, we calculate the constraints
in the $r_g$-$r_\gamma$ plane, coming from the limit on
$BR(B\to X_s\phi)$,
under the same assumptions made in refs. \cite{dh,dht}, which allow
a straightforward calculation of the branching ratio. In terms of
$r_g$-$r_\gamma$, the following expression is obtained:
\bea
BR(B\to X_s\phi)&=& a_1 \vert C_\phi (m_b)\vert^2+a_2 \alpha_s(m_b)
Re(C_\phi (m_b))r_g+a_3\alpha_e Re(C_\phi (m_b))r_\gamma\nonumber\\
&&+a_4 \alpha_s(m_b)^2r_g^2 +a_5 \alpha_e^2 r_\gamma^2+
a_6 \alpha_s(m_b)\alpha_e r_g r_\gamma,
\label{eq:brphi}
\eea
where the coefficients $a_i$ are functions of the masses $m_b$, $M_\phi$,
$m_s$. They are explicitly given in the Appendix. 

Beside $C_{11}$ and $C_{12}$, eq. (\ref{eq:brphi}) depends on the coefficient
$C_\phi$. It appears when the matrix elements
of the relevant penguin and electropenguin operators are evaluated using
the factorization method and it is given by \cite{dh}
\begin{eqnarray}
C_\phi (\mu)&=&C_3^{eff}(\mu)+C_4^{eff}(\mu)+C_5^{eff}(\mu)
+\xi\left(C_3^{eff}(\mu)+C_4^{eff}(\mu)+C_6^{eff}(\mu)\right)\nonumber\\
& &-\frac{1}{2}\left[C_7^{eff}(\mu)+C_9^{eff}(\mu)+C_{10}^{eff}(\mu)+\xi
\left(C_8^{eff}(\mu)+C_9^{eff}(\mu)+C_{10}^{eff}(\mu)\right)\right],
\label{eq:cphi}
\end{eqnarray}
where $\xi=1/N$. Here $N$ is the number of colours, taken
as a free parameter to account for possible deviations from the factorization
result.
Since the magnetic operators contribute to eq. (\ref{eq:brphi}) through
$O(\alpha_s)$ ($O(\alpha_e)$) matrix elements, it is mandatory to calculate
$C_\phi$ at NLO. 
The effective coefficients\footnote{Notice that the denomination ``effective
coefficient'' refers to completely different definitions in the case of
$C_3-C_{10}$ and $C_{11}-C_{12}$.}
$C_3^{eff}-C_{10}^{eff}$ are defined so that they include the NLO
contributions
coming from the matrix elements at $O(\alpha_s)$ ($O(\alpha_e)$),
calculated in perturbation theory \cite{dh,kps}. With such a definition,
$C_3^{eff}-C_{10}^{eff}$ become complex functions. We note that the
calculations of these coefficients in refs. \cite{dh,kps} are not complete.
A full NLO perturbative calculation of the matrix elements would
allow for a complete cancellation (at the considered order) of the scheme
dependence of the NLO Wilson coefficients that appear in $C_\phi$.
In our case, we use NLO coefficients $C_1-C_{10}$ calculated in the
't Hooft--Veltman scheme, including only the largest contribution to the NLO
matrix elements coming from $Q_2$, as in ref. \cite{dh}. We have estimated
the residual scheme dependence ($\sim 15\%$), by comparing our effective
coefficients to those given in ref. \cite{dh}. We consider it negligible, 
compared with the other uncertainties present in the calculation.

\begin{figure}[t]
\centering
\epsfxsize=0.9 \textwidth
\leavevmode\epsffile{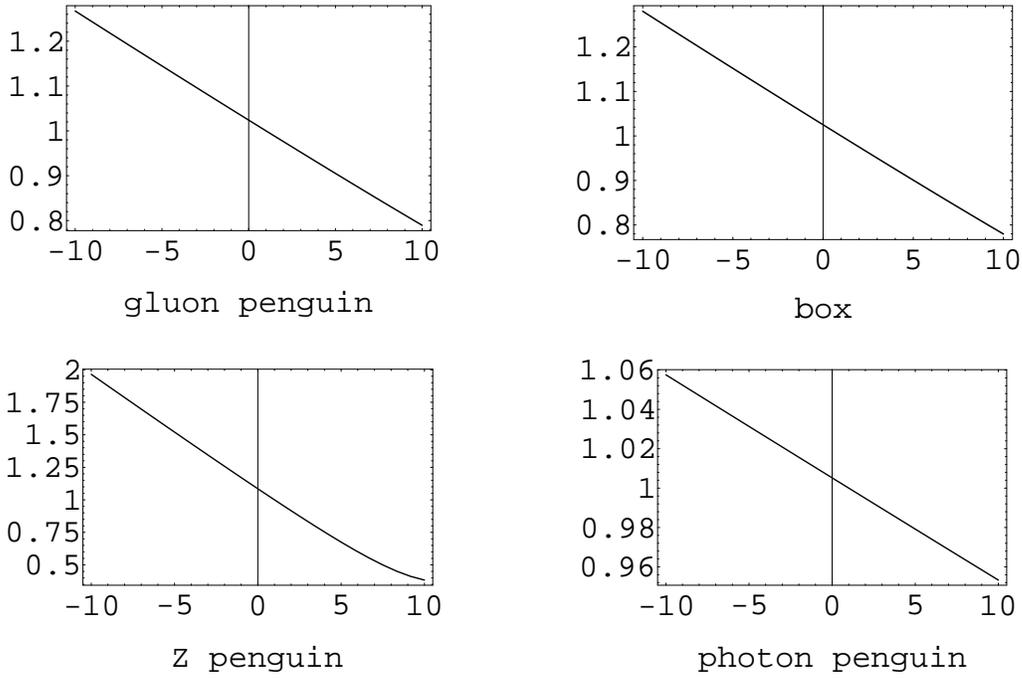}
\caption[]{\it{The coefficient $\vert C_\phi\vert$ as a function of
$C_3(M_W)$-$C_{10}(M_W)$, normalized to the Standard Model. The eight
initial conditions are parametrized in terms of the four contributions coming
from the box and the $g$, $Z$, $\gamma$ penguin diagrams.}}
\label{fig:inicond}
\end{figure}

The presence of $C_\phi$ in eq. (\ref{eq:brphi}) could make our
analysis much more involved, since this coefficient depends on several
initial conditions other
than $C_{11}(M_W)$ and $C_{12}(M_W)$. However $C_\phi$ is quite insensitive to
the initial conditions of the penguin and electropenguin operators $Q_3-Q_{10}$.
In fig. \ref{fig:inicond} we show $\vert C_\phi \vert$ as a function of these
initial conditions, parametrized in terms of
the contributions of the gluon, $Z$, $\gamma$ penguin and the
box diagrams, normalized to the Standard Model.
The reason of this weak dependence is twofold. On the one
hand, the largest terms in $C_\phi$ come from the gluon penguin operators
$Q_3-Q_6$,
which are insensitive to their initial conditions because the large
mixing with $Q_1-Q_2$ dominates their
renormalization group evolution down to $\mu=O(m_b)$.
On the other hand, the effective coefficients $C_i^{eff}$ get a large NLO
contribution from the $O(\alpha_s)$ matrix elements, which do
not depend on the initial conditions of the Wilson coefficients.
This implies that $BR(\bfi )$ in general gives a poor probe of new
physics effects, contrary to what is often stated in the literature.

Therefore, it is reasonable to assume $C_\phi$ as a constant in the analysis,
so that the $BR(B\to X_s\phi)$ could be used
to put constraints in the $r_g-r_\gamma$ plane. Unfortunately this
constraint suffers from large theoretical uncertainties. 
For example the predicted branching ratio changes by a factor of 2 as 
we vary $\xi$ between $1/2$ and $1/3$, which is a popular way to account for
the uncertainty of the factorization method \cite{dh}. Moreover the assumption
on the decay kinematics is not under control and the related uncertainty is
not quantifiable.
Just for illustrative purposes we show in fig.~\ref{fig:r11r12} the constraint
coming from eq.~(\ref{eq:brphi}), assuming $\xi=1/3$. For larger values of
$\xi$, the bound on $r_g$ from $B\to X_s\phi$ becomes more stringent, 
disfavouring even further the solutions of the $\bsl$--$n_c$ problems
corresponding to positive $r_g$. However
if we assign an overall uncertainty of a factor of 3 in the prediction of the
branching ratio, any significant bound on $r_g$ disappears.

An enhanced $b\to s g$ decay rate also affects the exclusive decays
$B\to K\pi$ \cite{dht}. We estimate $BR(B^-\to\bar K^0\pi^-)\simeq 10^{-5}
(1+0.1~r_g)^2$, which corresponds to an effect of order $1$ for $\vert r_g\vert
\sim 10$.
Again the theoretical uncertainty of this prediction is large and not
under control, so we prefer not to show this constraint in
fig.~\ref{fig:r11r12}.

We have seen how present constraints allow large enhancements of $BR(\bsg )$.
Measurements of $BR(\bsf )$ require, however, a precise correlation between 
$r_g$ and $r_\gamma$. The main difficulty to solve the $\bsl$--$n_c$
problem in terms of new physics is to explain this correlation. We now turn
to a discussion of models in which this is possible.

\section{An Illustrative Model}
\label{sec:model}

The suggestion that anomalously large $BR(\bsg )$ can explain a reduction
of $\bsl$ was first made in ref.~\cite{bh}, where it was assumed
that the new effective $bsg$ interaction is mediated by virtual charged
Higgs boson exchange. This possibility is now ruled out by a combination
of the constraints from $B\to X \tau \bar \nu$, $\bsf$, and $Z\to b\bar b$.
The main difficulty of models where the $bsg$ interaction arises from
charged-Higgs exchange, shared by most other models with weakly-interacting
new particles, is that generically $r_g\sim r_\gamma$. Constraints from
$BR(\bsf )$ then allow only small enhancements of $r_g$.

The possibility that flavour-changing quark--squark--gluino interactions
can generate large coefficients for the chromomagnetic operator was first
suggested in ref.~\cite{kag}. Now the loop generating $O_{11}$
has a large Casimir factor, which is not present in the loop generating
$O_{12}$, and one obtains $r_g/r_\gamma\sim$ 5--7. This allows a significant
enhancement of $\bsg$, especially if we consider solutions with negative
$r_\gamma$.

We have in mind the case in which the squark mass matrix is not diagonal
in the quark mass eigenbasis. This situation is generic in supersymmetric
models derived from supergravity with non-minimal K\"ahler metric. For
simplicity we consider separately two possibilities. In the first case,
we assume that flavour non-diagonal entries of the squark mass matrix
appear only in the left sector and there is no left--right squark mixing.
The new contributions to the Wilson coefficients $C_{11}$ and $C_{12}$,
evaluated at
the scale of supersymmetric particle masses, are \cite{bbm}
\beq
C_{11}=\frac{Z}{\mg^2}\sum_{i=1}^3U_{ib}^*U_{is}g_2(\mg^2/\tm_i^2)
\label{c8l}
\eeq
\beq
g_2(x)=\frac{x(-11x^2+40x+19)}{36(x-1)^3}+\frac{x^2(x-9)\log x}{6(x-1)^4}
\eeq
\beq
C_{12}=\frac{Z}{\mg^2}\sum_{i=1}^3U_{ib}^*U_{is}f_2(\mg^2/\tm_i^2)
\label{c7l}
\eeq
\beq
f_2(x)=\frac{2x(-2x^2-5x+1)}{27(x-1)^3}+\frac{4x^3\log x}{9(x-1)^4}
\eeq
\beq
Z\equiv \frac{\sqrt{2}\pi \alpha_s}{V_{tb}^*V_{ts}G_F}
\eeq
Here $\mg$ is the gluino mass and $\tm_i^2$ are the eigenvalues
of the down-squark squared-mass matrix, which is diagonalized by the
$3\times 3$ unitary matrix $U$. If $\tm_i$ are nearly degenerate,
eqs.~(\ref{c8l}) and (\ref{c7l}) can be Taylor-expanded around the
common squark mass $\tm$:
\beq
C_{11}=Z\frac{\delta \tm_{b_Ls_L}^2}{\tm^4}g_3(\mg^2/\tm^2)
\eeq
\beq
g_3(x)=\frac{(x^2+172x+19)}{36(x-1)^4}+\frac{x(x^2-15x-18)\log x}{6(x-1)^5}
\eeq
\beq
C_{12}=Z\frac{\delta \tm_{b_Ls_L}^2}{\tm^4}f_3(\mg^2/\tm^2)
\eeq
\beq
f_3(x)=\frac{2(-17x^2-8x+1)}{27(x-1)^4}+\frac{4x^2(x+3)\log x}{9(x-1)^5}
\eeq
Here $\delta \tm_{b_Ls_L}^2\equiv
\sum U_{ib}^*U_{is}\tm_i^2$ corresponds
to the flavour non-diagonal mass insertion.

In the second case we consider, the down-squark mass matrices for the left and
right sectors are both diagonal, but there are flavour non-diagonal
left--right (LR) 
mixing terms. Since these terms are not generated by supersymmetry 
breaking, in the limit of vanishing
Yukawa coupling constants, 
we will assume here that they are proportional to the corresponding Yukawa
coupling. In this case, the mass-insertion approximation is always
adequate and the new contributions to the Wilson coefficients $C_{11}$
and $C_{12}$ are
\beq
C_{11}=Z\frac{\delta \tm_{b_Rs_L}^2}{m_b\tm^3}g_4(\mg^2/\tm^2)
\eeq
\beq
g_4(x)=\sqrt{x}\left[
-\frac{2(x+11)}{3(x-1)^3}+\frac{(-x^2+16x+9)\log x}{3(x-1)^4} \right]
\eeq
\beq
C_{12}=Z\frac{\delta \tm_{b_Rs_L}^2}{m_b\tm^3}f_4(\mg^2/\tm^2)
\eeq
\beq
f_4(x)=\sqrt{x}\left[
\frac{4(5x+1)}{9(x-1)^3}-\frac{8x(x+2)\log x}{9(x-1)^4} \right].
\eeq
The two cases here considered are the simplest because they generate
only operators with the same chiral structure as in the Standard Model.
They also describe the generic features of more complicated squark mass
matrices.

By varying the relevant supersymmetric parameters under the requirement
that all squark masses are larger than 100 GeV and $\mg > 200$ GeV, 
we find that the new
interactions can generate values of $r_g$ and $r_\gamma$ within the 
regions illustrated in
fig.~\ref{fig:r11r12}.
The prediction has a strong correlation in the $r_g$--$r_\gamma$ plane
because the ratio $r_g/r_\gamma$ depends on the supersymmetric parameters
only weakly, through the loop functions. Thus, in spite of its generic features,
the model can make a rather precise prediction of this ratio.
It is interesting that it is indeed possible to reach a region in which
$\bsg$ has the correct enhancement to solve the $\bsl$--$n_c$ puzzle.
The solution always corresponds to the case in which the signs 
of both $C_{11}$ and $C_{12}$
are opposite to the Standard Model results. In the case of
left--left (LL) mixings this region is achieved when the mixing between the
$s$ and $b$ squark is maximal\footnote{Indeed the mixing is so large that
the mass-insertion approximation is not valid. The region in
fig.~\ref{fig:r11r12} was
obtained by using the complete expressions in eqs.~(\ref{c8l}) and 
(\ref{c7l}).} and when the lightest squark mass\footnote{Here we refer to
the mass of the lightest squark,
which can be considerably lighter than the other squarks, because of the
large mixing. Tevatron bounds on squark masses \cite{tev} do not directly
apply here, since they are obtained under the assumption of three families of
degenerate squarks.} is
around 100 GeV and $\mg$ around 200 GeV.
This means that supersymmetry could be soon discovered at the Tevatron,
but no light squark has to be expected at LEP2. Similar conclusions
apply to the case of LR mixing. However if the LR mixing
were not proportional to $m_b$, the same effects could be obtained for
much heavier squarks and gluinos, although the existence of colour-breaking
minima could impose significant constraints on the parameter space. 
Finally notice that, in the case of a
very light gluino ($\mg \sim 1$ GeV) the ratio $r_g/r_\gamma$ can become
much larger than that shown in fig.~\ref{fig:r11r12},
especially for LR mixings. In this
case it is possible to have $r_g$ as small as $-10$ for positive values of
$r_\gamma$, and almost reach the shaded region in fig.~2 corresponding to 
positive $r_\gamma$ and negative $r_g$.

Given a definite model of flavour violation, we can consider more tests on
its consistency than in the case of the model-independent analysis of
sec.~\ref{sec:modind}.
We now assume that gluino-mediated flavour violations occur only
between the second and third generation of the down quark--squark sector
and consider different processes with $|\Delta B|\ne 0$, which can be of 
experimental interest.

The $|\Delta B|=2$ transitions will affect the $B_s$--$\bar B_s$ mixing. For 
LL mass insertions, the new contribution to $\Delta m_{B_s}$ is \cite{bbm}
\beq
\Delta m_{B_s} =\frac{4}{3}B_{B_s}f_{B_s}^2m_{B_s}\frac{\alpha_s^2}{\tm^2}
\left(\frac{\delta \tm_{b_Ls_L}^2}{\tm^2}\right)^2 G(\mg^2/\tm^2)
\eeq
\beq
G(x)=\frac{1}{216}\left[ \frac{(-2x^3+27x^2+144x+11)}{(x-1)^4}-
\frac{2x(51x+39)\log x}{(x-1)^5}\right]\,.
\eeq
This contribution is much larger than the Standard Model 
one,
making the
discovery of $B_s$--$\bar B_s$ mixing even more arduous.
However, for $x\simeq 2.4$, the function $G(x)$ has a zero and
therefore, for particular values of the squark and gluino masses, we can
reduce the total contribution to $\Delta m_{B_s}$. Nevertheless, the
generic prediction of the model is that $B_s$--$\bar B_s$ mixing will
not be discovered soon.

The $|\Delta B|=1$ transitions can induce the flavour-changing decay $Z\to
\bar bs$. For LL transitions \cite{zbs}
\beq
BR(Z\to \bar b s)=\frac{\alpha \alpha_s^2m_Z}
{18\pi^2\sin^2\theta_W \cos^2\theta_W}\left( 1-\frac{2}{3}\sin^2\theta_W
\right)^2\left| U_{ib}^*U_{is}F(\tm_i^2,\mg^2,m_Z^2)\right|^2
\label{zap}
\eeq
\beq
F(\tm_i^2,\mg^2,m_Z^2)=\int_0^1 dx \int_0^{1-x}dy \log\left[ 
\frac{\tm_i^2 (x+y)+\mg^2(1-x-y)-m_Z^2xy}{\tm_i^2(1-x)+\mg^2x}\right]\,.
\eeq
Assuming the validity of the mass insertion approximation, eq.~(\ref{zap})
can be simply expressed in terms of $r_g$. For $\tm^2=\mg^2>m_Z^2$, we find
\beq
BR(Z\to \bar b s)=5\times 10^{-8}~ r_g^2\,.
\eeq
Unfortunately, observation of this decay mode represents a real experimental
challenge. LEP1 can at best reach a sensitivity on $BR(Z\to \bar b s)$ of
about $10^{-3}$ \cite{yk}.

Finally we consider FCNC decay processes of $B$ mesons, such as $B\to
X_s \nu \bar \nu$ and $B\to X_s \ell^+ \ell^-$. It is known \cite{bbm}
that gluino-mediated interactions do not affect the $Z$ penguin operator.
Indeed, the effective $b$-$s$-$Z$ vertex turns out to be proportional to the
$b$-$s$-$\gamma$ vertex. QED gauge invariance then implies that the $Z$ penguin
is suppressed by a factor $q^2/m_Z^2$, where $q^2$ is the momentum transfer.
Because of this suppression we do not expect new contributions to $B\to
X_s \nu \bar \nu$. On the other hand, the process $B\to X_s \ell^+ \ell^-$
receives new contributions from the gluino-mediated $\gamma$ penguin 
operator and from the electromagnetic dipole operator, which has here a
sign opposite to the Standard Model result. This change of sign 
implies important modifications both in the rate and the lepton asymmetries
of $B\to X_s \ell^+ \ell^-$ \cite{agm}. The new contribution to the
$\gamma$ penguin is however rather modest.
We find that LL transitions give
\beq
C_e=\frac{Z}{\mg^2}\sum_{i=1}^3U_{ib}^*U_{is}f(\mg^2/\tm_i^2)
\label{c9}
\eeq
\beq
f(x)=\frac{2x(11x^2-7x+2)}{81(x-1)^3}-\frac{12x^4\log x}{81(x-1)^4}\,,
\eeq
where $C_e$ is the Wilson coefficient at the weak scale of the operator
\beq
O_e=\frac{\alpha_e}{2\pi} (\bar s_\alpha b_\alpha)_{V-A} (\bar e e)_V\,.
\eeq
Using the mass-insertion approximation and taking for simplicity
$\mg^2/\tm_i^2=1$, eq.~(\ref{c9}) can be rewritten as
\beq
\frac{C_e}{C_e^{SM}}\simeq 5\times 10^{-2}~ r_g\,.
\eeq

\section{Conclusions}
In conclusion we have discussed how large $BR(\bsg )$ can reconcile the
predictions for $\bsl$ and $n_c$ with present experimental measurements.
Constraints from $BR(\bsf )$ require a precise correlation between new
physics effects in the coefficients of the chromomagnetic and electromagnetic
dipole operators. On the other hand $BR(\bfi )$ does not set any significant
constraints on the parameter space, mainly because of the large theoretical
uncertainties involved in the calculation. We have also shown that
$BR(\bfi )$ does not provide a good probe of new physics effects because
the relevant coefficient $C_\phi(m_b)$ is rather insensitive to the
initial conditions at the weak scale of the gluon--penguin operators.

Finally we have discussed how a supersymmetric model with gluino-mediated
flavour violations can account for a large value of $BR(\bsg )$, consistently
with all other constraints from $\Delta B=1$ and $\Delta B=2$ FCNC processes.
The theory has a rather precise correlation between the predictions for
$r_g$ and $r_\gamma$, which is fairly independent of the specific model
assumptions. Although the enhancement of $BR(\bsg )$ is achieved only for
particular values of the parameter space and it is evidently not a general
consequence of the model, it is interesting to know that supersymmetry can
potentially predict $BR(\bsg )\simeq 5$--$10\%$, and solve the $\bsl$--$n_c$
puzzle. Future experiments will certainly be able to test this scenario.

\section*{Note added}
After completion of this work, we received a note by
A. Kagan who informed us that their new results on $B\to X_s\phi$
agree with those presented in this paper \cite{kp}.

\section*{Acknowledgements}
We had useful discussions with A. Ali, P. Ball,
G. Martinelli and M. Neubert.
One of us, E.G., thanks the CERN Theory Division and the Department of Physics,
University of Southampton, for their kind hospitality during the completion
of this work.

\section*{Appendix}
In this appendix we report on the result of our calculation of
$BR(b\to s\phi)$. As in refs. \cite{dh,dht}, we
assume a pure two-body decay and use the factorization
method to estimate the relevant matrix elements.
We have computed both the contributions of $Q_{11}$ and $Q_{12}$.
The amplitude we obtain is
\bea
A(b\to s\phi)&=&
 -\frac{G_F}{\sqrt{2}}V_{tb}V_{ts}^* i g_\phi\epsilon_\mu
 \left\{2 C_\phi(m_b) \bs\gamma^\mu Lb
        +\frac{\alpha_s(m_b) m_b}{8\pi k^2}\frac{N^2-1}{2N^2} C_{11}(m_b)
         \left[m_b \bs\gamma^\mu Lb\right.\right.\nonumber\\
     && \left.\left.+\left(6+4\frac{m_s^2}{M_\phi^2}\right)p_b^\mu \bs Rb
        -m_s\left(1-2\frac{m_b^2+2M_\phi^2-m_s^2}{M_\phi^2}\right)
           \bs\gamma^\mu Rb
         -4\frac{m_b m_s}{M_\phi^2}p_b^\mu \bs L b\right]\right.\nonumber\\
     && \left.+\frac{\alpha_e m_b}{8\pi k^2} C_{12}(m_b)\left[9 m_b
        \bs\gamma^\mu Lb
        -\left(10-4\frac{m_s^2}{M_\phi^2}\right)p_b^\mu \bs Rb
         -m_s\left(1\right.\right.\right.\nonumber\\
     && \left.\left.\left.-2\frac{m_b^2+2M_\phi^2-m_s^2}{M_\phi^2}\right)
           \bs\gamma^\mu Rb
         -4\frac{m_b m_s}{M_\phi^2}p_b^\mu\bs Lb\right]
 \right\},
\label{eq:ampphi}
\eea
where $k^2=(m_b^2+m_s^2-M_\phi^2/2)/2$, $C_\phi(m_b)$ has been introduced in eq.
(\ref{eq:cphi}) and $g_\phi$ is defined by $\langle\phi\vert \bs\gamma^\mu
s\vert 0\rangle=ig_\phi\epsilon^\mu$.  This result
does not fully agree with those of refs. \cite{dh,dht}\footnote{Numerically our
branching ratio is $\sim 25\%$ smaller, including the effect of the
residual scheme dependence.}.

From eq. (\ref{eq:ampphi}), we obtain the branching ratio
\bea
BR(b&\to& s\phi)=\frac{\Gamma(b\to s\phi)}{\Gamma(b\to c e\nu)}
BR_{SL}^{exp}=
  \frac{\vert V_{tb}V_{ts}^*\vert^2}{\vert V_{cb}\vert^2} 
   \frac{6\pi^2 g_\phi^2 BR_{SL}^{exp}}{g(m_c/m_b)\Omega(m_c/m_b,m_b) m_b^4}
   \lambda(m_b^2,M_\phi^2,m_s^2)^{1/2}\nonumber\\
&&\left[\vert C_\phi(m_b)\vert^2\left( 2 + 2\frac{m_b^2}{M_\phi^2}
     - 4 \frac{M_\phi^2}{m_b^2}
     + 2 \frac{m_s^4}{m_b^2 M_\phi^2} + 2 \frac{m_s^2}{m_b^2}
     - 4 \frac{m_s^2}{M_\phi^2} \right)\right.\nonumber\\
&&+\frac{\alpha_s(m_b)}{8\pi} 
     \frac{2 \mbox{Re}\left(C_\phi(m_b)\right) C_{11}(m_b)}{k^2}
     \left(\frac{N^2-1}{2N^2}\right)\left(- 5 m_b^2
     -20 \frac{m_b^2 m_s^2}{M_\phi^2}
     + 4 \frac{m_b^4 }{M_\phi^2}+ M_\phi^2\right.\nonumber\\ 
&& \left.+ 16 \frac{m_s^4}{M_\phi^2} - 23 m_s^2 \right)
   +\frac{\alpha_e}{8\pi}\frac{2 \mbox{Re}\left(C_\phi(m_b)\right) C_{12}(m_b)}
       {k^2}\left(
      19 m_b^2  - 20 \frac{m_b^2 m_s^2}{M_\phi^2} + 4 \frac{m_b^4}{M_\phi^2}
      \right.\nonumber\\
&&\left.- 23 M_\phi^2 + 16 \frac{m_s^4}{M_\phi^2} + m_s^2 \right)
   +\left(\frac{\alpha_s(m_b) C_{11}(m_b)}{8\pi k^2}\right)^2
    \left(\frac{N^2-1}{2N^2}\right)^2\Biggl(
    \frac{31}{2} m_b^2 M_\phi^2\nonumber\\
&& - 4 \frac{m_b^2 m_s^6}{M_\phi^4} + 16 \frac{m_b^2 m_s^4}{M_\phi^2}
       - 54 m_b^2 m_s^2 - 19 m_b^4 -4 \frac{m_b^4 m_s^4}{M_\phi^4}
       + 4 \frac{m_b^6 m_s^2}{M_\phi^4}
       \nonumber\\
&&\left. + 8 \frac{m_b^6}{M_\phi^2} + \frac{15}{2} M_\phi^2 m_s^2
      -\frac{9}{2} M_\phi^4
      + 4 \frac{m_s^8}{M_\phi^4} - 8 \frac{m_s^6}{M_\phi^2} + m_s^4 \right)
      +\left(\frac{\alpha_e C_{12}(m_b)}{8\pi k^2}\right)^2
      \left( - \frac{177}{2} m_b^2 M_\phi^2
      \right.\nonumber\\
&&\left. - 4 \frac{m_b^2 m_s^6}{M_\phi^4}
     + 112 \frac{m_b^2 m_s^4}{M_\phi^2} - 62 m_b^2 m_s^2 + 93 m_b^4
     - 4 \frac{m_b^4 m_s^4}{M_\phi^4}- 96 \frac{m_b^4 m_s^2}{M_\phi^2}
      +4 \frac{m_b^6 m_s^2}{M_\phi^4}
      \right.\nonumber\\
&&\left.  + 8 \frac{m_b^6}{M_\phi^2}+ \frac{47}{2} M_\phi^2 m_s^2
      - \frac{25}{2} M_\phi^4 + 4 \frac{m_s^8}{M_\phi^4}
      - 8 \frac{m_s^6}{M_\phi^2} - 7 m_s^4 \right) \nonumber\\
&&+\frac{\alpha_s(m_b) C_{11}(m_b)}{8\pi k^2} \frac{\alpha_e C_{12}(m_b)}
      {8\pi k^2} \left(\frac{N^2-1}{2N^2}\right)\left(
     - 41 m_b^2 M_\phi^2 - 8 \frac{m_b^2 m_s^6}{M_\phi^4}
     + 192 \frac{m_b^2 m_s^4}{M_\phi^2}\right.\nonumber\\
&&\left. - 212 m_b^2 m_s^2 + 10 m_b^4
     -8 \frac{m_b^4 m_s^4}{M_\phi^4}-128 \frac{m_b^4 m_s^2}{M_\phi^2} +
      8 \frac{m_b^6 m_s^2}{M_\phi^4} + 16 \frac{m_b^6}{M_\phi^2}
      - 65 M_\phi^2 m_s^2\right.\nonumber\\
&&\left.\left.  + 15 M_\phi^4 + 8 \frac{m_s^8}{M_\phi^4}
      - 48 \frac{m_s^6}{M_\phi^2}+ 90 m_s^4\right)
      \right],
\label{eq:brphi1}
\eea
where $N$ is the number of colours, $\lambda(m_1,m_2,m_3)=
(1-m_2/m_1-m_3/m_1)^2-4 m_2m_3/m_1^2$,
$g(z)=1-8 z^2+8 z^6-z^8-24 z^4 \ln(z)$ is the phase-space correction
and $\Omega(z,\mu)\simeq 1-\frac{2\alpha_s(\mu)}{3\pi}\left[\left(\pi^2-
\frac{31}{4}\right)(1-z)^2+\frac{3}{2}\right]$ is the QCD
correction to the semileptonic decay rate.

The cofficients $a_i$ of eq. (\ref{eq:brphi}) can be immediately obtained
from eq. (\ref{eq:brphi1}).

\end{document}